\documentstyle[amssymb,prl,aps,epsfig]{revtex}

\begin{document}
\title{Implications of the isotope effects on the magnetization, magnetic
torque and susceptibility}
\author{T. Schneider}
\address{Physik-Institut der Universit\"{a}t Z\"{u}rich,\\
Winterthurerstrasse 190, CH-8057,Switzerland} \maketitle

\begin{abstract}
We analyze the magnetization, magnetic torque and susceptibility
data of La$_{2-x}$Sr$_{x}$Cu$^{16,18}$O$_{4}$ and YBa$_{2}$
$^{63,65}$Cu$_{3}$O$_{7-\delta }$ $\ $near $T_{c}$ in terms of the
universal 3D-XY scaling relations. It is shown that the isotope
effect on $T_{c}$ mirrors that on the anisotropy $\gamma $.
Invoking the generic behavior of $\gamma $ the doping dependence
of the isotope effects on the critical properties, including
$T_{c}$, correlation lengths and magnetic penetration depths are
traced back to a change of the mobile carrier concentration.

\end{abstract}


\bigskip

\begin{figure}[tbp]
\centering
\includegraphics[totalheight=6cm]{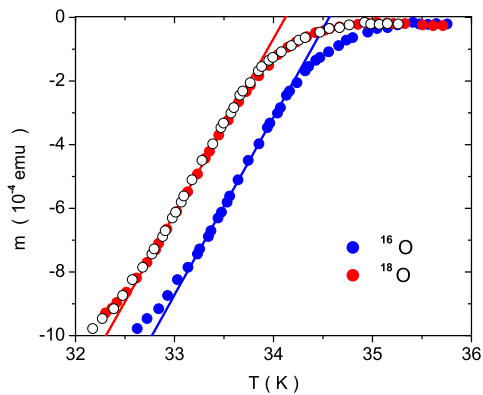}
\caption{$^{16,18}m\left( T\right) $ \emph{vs.} $T$ for a
La$_{1.85}$Sr$_{0.15} $Cu$^{16,18}$O$_{4}$ powder sample at $6.5$
Oe taken from Batlogg \emph{et al.}\protect\cite{batlogg}. The
parallel straight lines are linear extrapolations, yielding the
mean-field estimates $^{16}T_{c0}\approx 34.56$ K,
$^{18}T_{c0}\approx 34.12$ K and $\Delta T_{c0}/T_{c0}\approx
-0.012$. The black circles are the $^{16}m\left( T\right) $ data
rescaled according to Eq.(\ref{eq4}) with $a\simeq 0.986$.}
\label{fig1}
\end{figure}

A lot of measurements of the oxygen and copper isotope effect on
the magnetization, susceptibility, magnetic torque \emph{etc}.
have been performed on a variety of cuprate
superconductors\cite{batlogg,crawford,franck2,zhaocu,franck1,hofer214,zhao,hkw,khasanovr}.
To obtain estimates for the shift of the transition temperature
$T_{c}$ upon isotope exchange, the respective mean-field
models\cite{abrikosov,kogan} have been invoked. For the
magnetization $m$ they imply that the mean-field transition
temperature $T_{c0}$ can be estimated by extrapolating $m\left(
T\right) $ linearly. To illustrate this traditional treatment we
displayed in Fig.\ref{fig1} the data of Batlogg \emph{et
al.}\cite{batlogg} for a nearly optimally doped
La$_{1.85}$Sr$_{0.15}$Cu$^{16,18}$O$_{4}$ powder sample in terms
of $^{16,18}m\left( T\right) $ \emph{vs.} $T$. The straight lines
are parallel linear extrapolations, yielding the mean-field
estimates $^{16}T_{c0}\approx 34.56$ K, $^{18}T_{c0}\approx 34.12$
K and $\Delta T_{c0}/T_{c0}\approx -0.012$, where $\Delta Y=\left(
^{18}Y-^{16}Y\right) /^{18}Y$. However, the reliability of these
estimates is doubtful because the dominant role of critical
fluctuations near $T_{c}$ \cite
{fisher,tshk,hubbard,x,babic,hoferdis,hofer,book,parks,tsdc} and
the isotope effects on the penetration
depth\cite{tshkprl,tsiso,tsrkhk,khasanov123f} \textit{etc. }are
neglected, the anisotropy is not taken into account, and the
choice of a linear and parallel portion appears arbitrary.

Noting that the isotope effects in cuprate superconductors pose a
fundamental challenge in its understanding, the shortcomings of
the traditional interpretation of magnetization and related data
call for a treatment that goes beyond mean-field and takes the
anisotropy into account. In this work we concentrate on the
critical regime of anisotropic extreme type II superconductors
where 3D-XY fluctuations dominate\cite
{fisher,tshk,hubbard,x,babic,hoferdis,hofer,book,parks,tsdc}.
Invoking the universal scaling functions for magnetization and
magnetic torque we analyze the magnetization data of
La$_{1.85}$Sr$_{0.15}$Cu$^{16,18}$O$_{4}$\cite {batlogg}, magnetic
torque data of
La$_{1.92}$Sr$_{0.08}$Cu$^{16,18}$O$_{4}$\cite{hofer214} and the
susceptibility data of YBa$_{2}$ $^{63,65}$Cu$_{3}$O$_{7-\delta }$
$\cite{zhaocu}\ $near $T_{c}$. An essential additional relation
emerges from the observation that data taken at fixed magnetic
field and on samples with $^{16}$O or $^{18}$O, as well as with
$^{63}$Cu or $^{65}$Cu, collapse near criticality within
experimental error, when the temperature is rescaled
appropriately. Although this property provides retrospectively
partial support for the traditional approach\cite
{batlogg,crawford,franck2,zhaocu,franck1,hofer214,zhao,hkw,khasanovr},
3D-XY scaling uncovers the essential role of the anisotropy
$\gamma $. Indeed, the change of $T_{c}$ is found to mirror the
shift of the anisotropy $\gamma $. As a consequence, the generic
shift of the temperature dependent magnetization, susceptibility
and magnetic torque upon isotope exchange at fixed magnetic field
does not provide estimates for the change of the transition
temperature only, as hitherto assumed\cite
{batlogg,crawford,franck2,zhaocu,franck1,hofer214,zhao,hkw,khasanovr}.
Together with the generic behavior of the anisotropy\cite
{hofer,suzuki,nakamura,willemin,kimura,sasagawa,tsw}, the doping
dependence of the isotope effects are then traced back to the
change $\Delta x_{u}$ of the underdoped limit $x_{u}$. It implies
a shift of the phase diagram in the temperature-doping plane
towards a slightly higher dopant concentration $x$, along with a
reduction of the mobile charge carrier concentration
$\overline{\delta }=x-x_{u}$. This contribution leads to a
negative shift of $T_{c}$. We identify a positive shift as well.
It stems from the change $\Delta \gamma _{0}$ of the critical
amplitude $\gamma _{0}$ at the quantum superconductor to insulator
transition. The magnitude and proportion of these contributions is
controlled by $\Delta x_{u}$ and $\Delta \gamma _{0}$. Their
values are material dependent. In any case, they control the
isotope effects and remain to be understood microscopically.
However, the emerging essential role of the anisotropy represents
a serious problem for two dimensional models as candidates to
explain superconductivity in the cuprates, and serves as a
constraint on future work towards a complete understanding.

Whenever 3D-XY fluctuations dominate the magnetization $m$ adopts
the scaling
form\cite{tshk,hubbard,x,babic,hoferdis,hofer,book,parks,tsdc}
\begin{equation}
\frac{m\left( T,\delta ,H\right)
}{T\sqrt{H}}=-\frac{k_{B}Q_{3}}{\Phi _{0}^{3/2}}\gamma \epsilon
^{3/2}\left( \delta \right) \frac{1}{\sqrt{z}}\frac{dG\left(
z\right) }{dz},\text{ \ }z=\frac{H\xi _{ab}^{2}}{\Phi
_{0}}\epsilon \left( \delta \right) ,  \label{eq1}
\end{equation}
where $\epsilon \left( \delta \right) =\left( \cos ^{2}\left(
\delta \right) +\sin ^{2}\left( \delta \right) /\gamma ^{2}\right)
^{1/2}$ and $\gamma =\xi _{ab}/\xi _{c}$ denotes the anisotropy.
$Q_{3}$ is a universal constant, $G\left( z\right) $ a universal
function of its argument, $\xi _{ab,c}$ the correlation lengths in
the $ab$-plane and along the $c$-axis, $H$ the magnetic field and
$\Phi _{0}$ the flux quantum. Close to the zero field transition
temperature $T_{c}$ the correlation lengths diverge as $\xi
_{ab,c}=\xi _{ab0,c0}\left| t\right| ^{-\nu }$ where $\nu \simeq
2/3$ and $t=T/T_{c}-1$. An essential implication is that in the
plot $m\left( T,\delta ,H\right) /\left( \gamma \epsilon
^{3/2}\left( \delta \right) T\sqrt{H}\right) $ \textit{vs.
}$z=\left( H\xi _{ab0}^{2}\epsilon \left( \delta \right) /\Phi
_{0}\right) \left| t\right| ^{-2\nu }$ or $x=z^{-1/2\nu }=\left(
\Phi _{0}/\left( H\xi _{ab0}^{2}\epsilon \left( \delta \right)
\right) \right) ^{1/2\nu }t$ the data fall close to criticality on
a single curve. For YBa$_{2}$Cu$_{3}$O$_{7-\delta }$ this scaling
property is experimentally well confirmed\cite{hubbard,babic}.
Because the magnetization exists at $T_{c}$ the combination
$m\left( T,\delta ,H\right) /\left( \gamma \epsilon ^{3/2}\left(
\delta \right) T\sqrt{H}\right) $ adopts the universal value
\cite{x,hoferdis,hofer,book,parks}
\begin{equation}
\frac{m\left( T_{c}\right) }{\gamma \left( T_{c}\right) \epsilon
^{3/2}\left( \delta \right)
T_{c}\sqrt{H}}=-\frac{k_{B}Q_{3}c_{3,\infty }}{\Phi _{0}^{3/2}},
\label{eq2}
\end{equation}
where $c_{3,\infty }$ is a universal constant\cite
{x,hoferdis,hofer,book,parks}. Thus, plotting $m\left( T\right)
/\left( \gamma \epsilon ^{3/2}\left( \delta \right)
\sqrt{H}\right) $ \textit{vs. }$T $, the data taken in different
fields cross at $T_{c}$. In powder samples and sufficiently large
anisotropy $\left( \gamma >>1\right) $ this relation reduces to
\begin{equation}
\frac{m\left( T_{c}\right) }{\gamma \left( T_{c}\right)
T_{c}\sqrt{H}}=-\frac{\pi k_{B}Q_{3}c_{3,\infty }\left\langle
\left| \cos \left( \delta \right) \right| ^{3/2}\right\rangle
}{2\Phi _{0}^{3/2}}. \label{eq3}
\end{equation}
As the oxygen isotope effect on the magnetization at fixed
magnetic field is concerned it implies that the relative shifts of
magnetization $m$, anisotropy $\gamma $ and $T_{c}$ are not
independent but related by $\Delta m\left( T_{c}\right) /m\left(
T_{c}\right) +\Delta \gamma \left( T_{c}\right) /\gamma \left(
T_{c}\right) +\Delta T_{c}/\Delta T_{c}=0$. On that condition it
appears impossible to extract $T_{c}$ and $\Delta T_{c}$ from the
temperature dependence of the magnetization taken in one
particular magnetic field. However, there is the special case
where close to criticality the data scales as
\begin{equation}
^{18}m\left( T\right) =^{16}m\left( aT\right) ,  \label{eq4}
\end{equation}
within experimental error. Subsequently it would also justify the
traditional method of extracting $\Delta
T_{c}/T_{c}$\cite{batlogg,crawford,franck2,zhaocu,franck1,hofer214,zhao,hkw,khasanovr}.
A glance to Fig.\ref{fig1} shows that this scale transformation is
well confirmed within experimental error, roughly given by the
size of the symbols of the data points. It yields $\Delta m\left(
T_{c}\right) /m\left( T_{c}\right) \simeq 0 $ with $a=$
$^{18}T_{c}/^{16}T_{c}\simeq 0.986$. Thus, $\Delta
T_{c}/T_{c}\simeq -\Delta \gamma \left( T_{c}\right) /\gamma
\left( T_{c}\right) \simeq -0.014$ for
La$_{1.85}$Sr$_{0.15}$Cu$^{16,18}$O$_{4}$. Although $\Delta
T_{c}/T_{c}\simeq -0.014$ is close to the traditional estimate ,\
$\Delta T_{c0}/T_{c0}\approx -0.012$, its significance is
fundamentally different. Indeed, because $\Delta m\left(
T_{c}\right) /m\left( T_{c}\right) \simeq 0$, the universal
relation $\Delta m\left( T_{c}\right) /m\left( T_{c}\right)
+\Delta \gamma \left( T_{c}\right) /\gamma \left( T_{c}\right)
+\Delta T_{c}/\Delta T_{c}=0$ reduces to
\begin{equation}
-\frac{\Delta T_{c}}{T_{c}}=\frac{\Delta \gamma \left(
T_{c}\right) }{\gamma \left( T_{c}\right) }=\frac{\Delta \xi
_{ab0}}{\xi _{ab0}}-\frac{\Delta \xi _{c0}}{\xi _{c0}}=1-a.
\label{eq5}
\end{equation}
Hence, the isotope effect on $T_{c}$ mirrors that on the
anisotropy and the critical amplitudes of the correlation lengths.
In virtue of the universal scaling expression (\ref{eq2}),
relation (\ref{eq4}) is obtained when the anisotropy $\gamma $
scales near $T_{c}$ as $\gamma \left( T\right) =a\gamma \left(
aT\right) $. The rescaled curves should then cross at $T_{c}$,
provided that the experimental uncertainties do not mask the
isotope induced change of $\xi _{ab0}$, the critical amplitude of
the in-plane correlation length. Considering the data shown in
Fig.\ref{fig1} we observe that the two curves coincide in the
critical regime within experimental error, roughly given by the
size of the symbols of the data points. Apparently, the resolution
of the crossing point requires considerably more accurate data.
Otherwise, as in the present case, the coincidence of the rescaled
data confirms the consistency with 3D-XY critical behavior, allows
to determine the rescaling factor $a$ around $T_{c}$, and with
Eq.(\ref{eq5}) to estimate the shifts $\Delta T_{c}/T_{c}$ and
$\Delta \gamma \left( T_{c}\right) /\gamma \left( T_{c}\right) $
rather accurately.

\begin{figure}[tbp]
\centering
\includegraphics[totalheight=6cm]{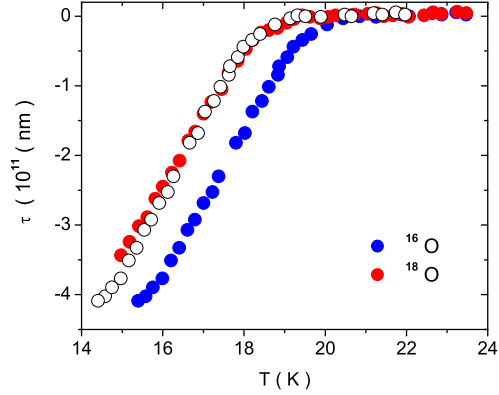}
\caption{Reversible magnetic torque $\tau \left( \delta \right) $
\textit{vs.} $T $ at $H=0.1$ T and $\delta = 45^{\circ} $ for
La$_{1.92}$Sr$_{0.08}$Cu$^{16,18}$O$_{4}$ taken from Hofer
\emph{et al.}\protect\cite{hofer214}. The black circles are
$^{18}m\left( T\right) \simeq ^ {16}m\left( aT\right) $ with
$a=0.936$.} \label{fig2}
\end{figure}

Before turning to the implications of these results, revealing
that the isotope effect on $T_{c}$ mirrors that on the anisotropy
$\gamma $, it is essential to explore the effect of the dopant
concentration. Since sufficiently dense data appears to be
available for underdoped La$_{1.92}$Sr$_{0.08}$Cu$^{16,18}$O$_{4}$
only, we are left with the reversible magnetic torque data of
Hofer \emph{et al.} \cite{hofer214} shown in Fig. \ref{fig2} in
terms of $\tau $ \emph{vs.} $T$. At $T_{c}$, fixed orientation and
magnitude of the applied field $\tau $ scales as $\tau \left(
T_{c}\right) =-constT_{c}\gamma \left( T_{c}\right) H^{3/2}$ \cite
{x,hoferdis,hofer,book,parks} and at fixed magnetic field the
shifts are related by $\Delta \tau \left( T_{c}\right) /\tau
\left( T_{c}\right) +\Delta \gamma \left( T_{c}\right) /\gamma
\left( T_{c}\right) +\Delta T_{c}/\Delta T_{c}=0$. From
Fig.\ref{fig2} it is seen that with $^{18}\tau \left( T\right)
=^{16}\tau \left( aT\right) $ and $a=0.936$ near coincidence is
achieved within experimental accuracy. Since the universal scaling
law for the magnetic torque\cite{x,hoferdis,hofer,book} is
essentially analogous to Eq.(\ref{eq2}) \ the near coincidence
implies $\Delta \tau \left( T_{c}\right) /\tau \left( T_{c}\right)
\simeq 0$ and\ (\ref{eq5}) holds in this case as well, so that
$\Delta T_{c}/T_{c}\simeq -\Delta \gamma \left( T_{c}\right)
/\gamma \left( T_{c}\right) \simeq -0.07$.

To check the generic significance of this scenario we also
analyzed the susceptibility data for the copper isotope effect in
YBa$_{2}$ $^{63,65}$Cu$_{3}$O$_{7-\delta }$ of Zhao \emph{et
al}.\cite{zhaocu}. For all four dopant concentrations, extending
from the underdoped to the optimally doped regime, we find that
$^{65}\chi \left( T\right) =$ $^{63}\chi \left( aT\right) $ is
satisfied within experimental error, as discussed below. Since
$\chi =m/H$ this strongly suggests that the scaling relation
(\ref{eq5}) holds for both, copper and oxygen isotope exchange,
irrespective of the doping level.

\begin{figure}[tbp]
\centering
\includegraphics[totalheight=6cm]{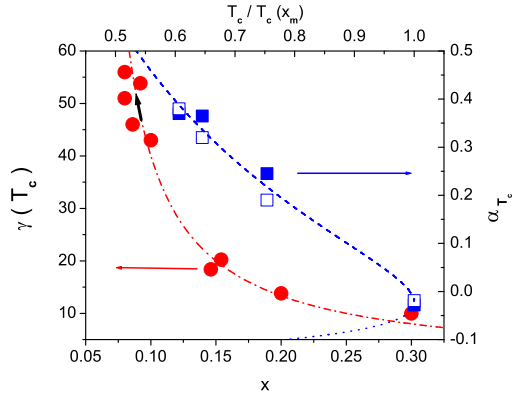}
\caption{ $\gamma \left( T_{c}\right) $ \textit{vs. }$x$ for
La$_{2-x}$Sr$_{x}$CuO$_{4}$ taken
from\protect\cite{hofer,suzuki,nakamura,willemin,kimura,sasagawa}
($\bullet $) and $\alpha _{T_{c}}$ \textit{vs. }$T_{c}/T_{c}\left(
x_{m}\right) $ for YBa$_{2}$ $^{63,65}$Cu$_{3}$O$_{7-\delta }$
taken from Zhao \emph{et
al.}\protect\cite{zhaocu}($\blacksquare$). The dash-dot line is
Eq.(\ref{eq7}) with $x_{u}=0.05$ and $\gamma _{0}=2$. The arrow
indicates the flow to the superconductor to insulator transition.
The open squares result from the scaling analysis of the
susceptibility data for the samples with $7-\delta =6.94$, $6.75$,
$6.63$ and $6.48$ in terms of $^{65}\chi \left( T\right) =$
$^{63}\chi \left( aT\right) $ yielding $a\simeq 1.0006$, $0.994$,
$0.990$ and $0.988$, respectively. The dashed and dotted curves
result from Eqs.(\ref{eq8}) and (\ref{eq9}) with $\Delta \gamma
_{0}/\gamma _{0}=-0.0082$, $\Delta x_{u}=0.0012$ and $T_{c}\left(
x_{m}\right) =92.37$ K. } \label{fig3}
\end{figure}

Having established the consistency with 3D-XY critical behavior,
together with the experimental facts that at $T_{c}$ and fixed
magnetic field $\Delta m/m$, $\Delta \tau /\tau $ and $\Delta \chi
/\chi $ vanish within experimental error, the isotope effect on
$T_{c}$ does not mirror that of the anisotropy only
(Eq.(\ref{eq5})), but is also subject to the other universal
relations of the 3D-XY universality class. In particular, $T_{c}$,
$\xi _{c0}$ and $\lambda _{ab0}$ are not independent but related
by the universal relation
\cite{tshk,x,book,parks,tsiso,peliasetto} $T_{c}=B\xi
_{c0}/\lambda _{ab0}^{2}$, where $B$ is a universal constant and
$\lambda _{ab0}$ the critical amplitude of the in-plane
penetration depth. This leads for the respective relative shifts
upon isotope exchange to the additional relation $\Delta
T_{c}/T_{c}=\Delta \xi _{c0}/\xi _{c0}-2\Delta \lambda
_{ab0}/\lambda _{ab0}$. The lesson is, whenever 3D-XY fluctuations
dominate, the isotope effects, e.g. on $T_{c}$,$\gamma $ and
$\lambda _{ab0}$ are not independent. These relations are
particularly useful to open a door towards the understanding of
the common origin of these effects. For example, given a generic
relationship between anisotropy $\gamma $ and mobile carrier
concentration $\widetilde{\delta }=x-x_{u}$ at fixed $x$, where
$x_{u}$, is the underdoped limit, the isotope effects in the
cuprates would arise from a shift of $x_{u}$ and the associated
change of $\widetilde{\delta }$. As shown in Fig.\ref{fig3} for
La$_{2-x}$Sr$_{x}$CuO$_{4}$, the generic doping dependence of
$\gamma $ is well established in a rich variety of cuprates in
terms of the empirical relation\cite{parks,tsw,tshknjp}
\begin{equation}
\gamma \left( T_{c}\right) =\frac{\gamma _{0}}{\widetilde{\delta
}}=\frac{\gamma _{0}}{x-x_{u}},  \label{eq7}
\end{equation}
where $\gamma _{0}$ is material dependent constant. Approaching
the underdoped limit, where the cuprates correspond to an
independent stack of sheets with thickness
$d_{s}$\cite{parks,tsw,tshknjp}, this relation follows from the
doping dependence of the critical amplitudes of the correlation
lengths. Since $\xi _{c0}$ tends to $d_{s}$, while $\xi _{ab0}$
diverges as $\xi _{ab0}=\overline{\xi _{ab0}}/\left(
x-x_{u}\right) $\cite {parks,tsw,tshknjp} we obtain $\gamma
_{0}=d_{s}/\overline{\xi _{ab0}}$, which is the critical amplitude
of the anisotropy at the quantum superconductor to insulator
transition at $x_{u}$. The doping dependence of the relative
isotope shifts is then traced back to a change of $\gamma _{0}$
and the shift $\Delta x_{u}$ of the underdoped limit and with that
to a change of the mobile carrier concentration $\widetilde{\delta
}=x-x_{u}$ according to
\begin{equation}
\frac{\Delta \gamma \left( T_{c}\right) }{\gamma \left(
T_{c}\right) }=\frac{\Delta \gamma _{0}}{\gamma _{0}}+\frac{\Delta
x_{u}}{\widetilde{\delta }}=\frac{\Delta \gamma _{0}}{\gamma
_{0}}+\frac{\Delta x_{u}}{x_{m}\left( 1\pm
\sqrt{1-T_{c}/T_{c}\left( x_{m}\right) }\right) },  \label{eq8}
\end{equation}
where we invoked the empirical relation between the hole
concentration $x$ and $T_{c}$ due to Presland \emph{et
al}.\cite{presland}. $x_{m}\simeq 0.16$ denotes optimum doping.
This leads to the important conclusion that the doping dependence
of the isotope effects in the cuprates stem from the shift of the
underdoped limit. Finally, combined with Eq.(\ref{eq5}) we obtain
\begin{equation}
\frac{\Delta \gamma \left( T_{c}\right) }{\gamma \left(
T_{c}\right) }=\frac{\Delta \gamma _{0}}{\gamma _{0}}+\frac{\Delta
x_{u}}{\widetilde{\delta }}=1-a=-\frac{\Delta
T_{c}}{T_{c}}=\frac{\Delta \xi _{ab0}}{\xi _{ab0}}-\frac{\Delta
\xi _{c0}}{\xi _{c0}}=-\frac{\Delta \xi _{c0}}{\xi
_{c0}}+\frac{2\Delta \lambda _{ab0}}{\lambda _{ab0}},  \label{eq9}
\end{equation}
relating the various relative shifts to the scaling factor $a$.
With our estimates $\Delta T_{c}/T_{c}\simeq -0.014$ ($x=0.15$)
and $\Delta T_{c}/T_{c}\simeq -0.07$ ($x=0.08$) for
La$_{2-x}$Sr$_{x}$CuO$_{4}$ and relations (\ref{eq8}) and
(\ref{eq9}) we obtain with $x_{u}=0.05$ for the essential, but
material dependent parameters, determining the doping and $T_{c}$
dependence of $\alpha _{T_{c}}=-\left( M/\Delta M\right) \Delta
T_{c}/T_{c}$ the values $\Delta \gamma _{0}/\gamma _{0}\simeq
-0.01$ and $\Delta x_{u}\simeq 0.0024$. To illustrate this feature
and to check the generic significance of this scenario further we
consider the copper isotope effect on $T_{c}$ in
YBa$_{2}$Cu$_{3}$O$_{7-\delta }$. As aforementioned, our scaling
analysis of the susceptibility data of Zhao \emph{et al}.\cite
{zhaocu} reveals full consistency with  $^{65}\chi \left( T\right)
=$ $^{63}\chi \left( aT\right) $ for all doping concentrations
within experimental error. Since $\chi =m/H$ the implications are
equivalent to those derived for the magnetization and
Eq.(\ref{eq9}) applies as well. The resulting estimates for
$\alpha _{T_{c}}=-(M/\Delta M)\Delta T_{c}/T_{c}$ are included in
Fig.\ref{fig3} and compared with those obtained from the
traditional extrapolation approach \cite{zhaocu}.More importantly,
given $\Delta \gamma _{0}/\gamma _{0}$ and $\Delta x_{u}$ the
$T_{c}$ dependence of $\alpha _{T_{c}}$ is readily calculated with
the aid of Eqs.(\ref{eq8}) and (\ref{eq9}). As shown in
Fig.\ref{fig3} in terms of the dashed line, agreement is achieved
with $\Delta \gamma _{0}/\gamma _{0}\simeq -0.0082$ and $\Delta
x_{u}\simeq 0.0012$. In comparison with Y$_{1-y}\Pr_{y}$Ba$_{2}$%
Cu$_{3}$ $^{16,18}$O$_{7-\delta }$ the data of Franck \emph{et
al.}\cite {franckpr} yields $\Delta \gamma _{0}/\gamma _{0}\simeq
-0.0060$ and $\Delta x_{u}\simeq 0.0019$.

In summary we have seen that near $T_{c}$, where 3D-XY
fluctuations are essential, the isotope effects on various
critical properties are not independent but related by universal
relations. Together with the observation, that data taken at fixed
magnetic field and on samples with $^{16}$O or $^{18}$O, as well
as with $^{63}$Cu or $^{65}$Cu, collapse near criticality within
experimental error, when the temperature is rescaled
appropriately, we derived an additional relationship. It reveals
the essential relevance of the anisotropy $\gamma $. Indeed, the
relative shift of $T_{c}$ was found to mirror that of the
anisotropy $\gamma $. As a consequence, the temperature shift of
the magnetization, susceptibility and the magnetic torque at fixed
magnetic field does not provide estimates for the change of the
transition temperature only, as hitherto assumed\cite
{batlogg,crawford,franck2,zhaocu,franck1,hofer214,zhao,hkw,khasanovr}.
Together with the generic behavior of the anisotropy\cite
{hofer,suzuki,nakamura,willemin,kimura,sasagawa}, the doping
dependence of the isotope effects was traced back to a change of
the underdoped limit $\Delta x_{u}$, or in other words, to a shift
of the phase diagram in the temperature-doping plane towards a
slightly higher dopant concentration, along with a reduction of
the mobile charge carrier concentration. This contribution leads
to a negative shift of $T_{c}$. We identified a positive shift as
well. It stems from the change of the critical amplitude of the
anisotropy $\gamma _{0}$ at the quantum superconductor to
insulator transition. The magnitude and proportion of these
contributions is controlled by $\Delta x_{u}$ and $\Delta \gamma
_{0}$. Their values turned out to be material dependent. In any
case, they control the isotope effects and remain to be understood
microscopically. However, the emerging essential role of the
anisotropy represents a serious problem for two dimensional models
as candidates to explain superconductivity in the cuprates, and
serves as a constraint on future work towards a complete
understanding. In addition, isotope exchange leads unavoidably to
lattice distortions. Their coupling with the in-plane penetration
depth was recently established\cite{tsrkhk}.

\acknowledgments The author is grateful to S. Kohout and J. Roos
for useful comments and suggestions on the subject matter.

\end{document}